\documentclass[]{aa}
\bibpunct{(}{)}{;}{a}{}{,}
\usepackage[varg]{txfonts}
\usepackage{graphicx}
\usepackage{upgreek}

\renewcommand{\pi}{\uppi}

\newcommand{\mdot}{\ensuremath{\dot{m}}}
\newcommand{\Msun}{\ensuremath{\rm M_\sun}}
\newcommand{\Msunyr}{\ensuremath{\rm M_\sun\, \rm yr^{-1}}}

\newcommand{\Gpc}{\ensuremath{\rm Gpc}}

\newcommand{\pc}{\ensuremath{\rm pc}}
\newcommand{\cmc}{\ensuremath{\rm cm^{-3}}}
\newcommand{\cmsq}{\ensuremath{\rm cm^{-2}}}
\newcommand{\Ry}{\ensuremath{\rm Ry}}
\newcommand{\AAA}{\ensuremath{{\rm \AA}}}

\renewcommand{\div}{\ensuremath{-}}

\newcommand{\qj}{Q~J0158-4325}

\begin{document}

\title{Apparent quasar disc sizes in the ``bird's nest'' paradigm
}


\author{P. Abolmasov}

\offprints{Pavel Abolmasov, \email{pavel.abolmasov@gmail.com}}

\institute{Tuorla Observatory, University of Turku, V\"ais\"al\"antie 20,
  FI-21500 Piikki\"o, Finland 
\and
Sternberg Astronomical Institute, Moscow State University,
  Universitetsky pr., 13, Moscow 119992, Russia,\\
\email{pavel.abolmasov@gmail.com}
}

\date{}

\abstract {Quasar microlensing effects make it possible to measure the accretion disc
sizes around distant supermassive black holes that are still well beyond the
spatial resolution of contemporary instrumentation. The sizes measured with
this technique appear inconsistent with the standard accretion disc
model. Not only are the measured accretion disc sizes larger, but their
dependence on wavelength is in most cases completely different from the
predictions of the standard model. } 
{We suggest that these discrepancies may arise not from non-standard accretion
  disc structure or systematic errors, as it was proposed before, but rather
  from scattering and reprocession of the radiation of the disc. 
In particular, the matter falling from the
gaseous torus and presumably feeding the accretion disc
may at certain distances become ionized and produce an extended
halo that is free from colour gradients.} {A simple analytical model is proposed
  assuming that a geometrically thick translucent inflow acts as a scattering
  mirror changing the apparent
  spatial properties of the disc. 
This inflow may be also identified with the broad line
  region or its inner parts.} {Such a model is able to explain the basic
properties of the apparent disc sizes, primarily their large values and their 
shallow dependence on wavelength. The only condition required is to scatter
a significant portion of the luminosity of the disc. This can easily be fulfilled
if the scattering inflow has a large geometrical thickness and clumpy
structure.} {}

\keywords{accretion, accretion discs -- quasars: general -- scattering -- 
  gravitational lensing: micro} 
\maketitle 

\section{Introduction}


While the mechanisms responsible for the spectral energy distributions of
bright active galactic nuclei are more or less understood in the framework of
standard disc accretion theory, there are certain observational
results that lack unambiguous interpretation. One of the most puzzling is the
behaviour of the accretion disc sizes measured through microlensing
effects. \citet{pooley07} have found the observed disc sizes to exceed the
standard theory predictions by one-two orders of magnitude. Later, in
\citet{morgan10}, this result was qualitatively confirmed, but the real value
of discrepancy appeared to be smaller -- about a factor of several. Even more
important is that the wavelength dependence of the radius seems to deviate
generally from the expected standard theory scaling. 
For a multi-temperature disc with a power-law temperature
dependence on radius, $T\propto R^{-p}$, the apparent radius should scale as
$R\propto \lambda^{1/p}$. In particular, for a standard disc,  $T \propto
R^{-3/4}$ \citep{SS73} and $R\propto \lambda^{4/3}$. 
\citet{blackburne12} find that for the larger
part of their 12-object sample, the apparent disc sizes in the UV/optical
range (0.1$\div$1$\rm \mu m$) show much weaker dependence on wavelength.
An evident
solution is to assume steeper temperature profile, but there seem to be no
physically motivated models able to reproduce the very small temperature slopes
$p> 1$ (see also discussion in \citet{floyd09}). Another solution is to
assume that the measured spatial scales are not representative of the
accretion disc itself. 
For instance, \citet{yan14} suggest that a sub-parsec black hole
binary model may reproduce some of the observational results. In this case,
the observed size of the emitting region is close to the separation of the
binary that does not depend on wavelength. 

It seems that this large-radius problem is typical for lower mass black
holes ($M \lesssim 10^9\Msun$), while the discs of the most massive
supermassive black holes better conform to the standard disc theory (see
Fig.~4 in \citet{superpaper}). 
Because the sample of lensed quasars is
biased towards brighter objects, one can suggest that smaller black holes
should have higher Eddington ratios, i.e. higher mass accretion rates in
Eddington units. Accretion upon smaller black holes should probably be even
super-Eddington \citep{SS73}. Super-Eddington mass accretion is expected to
lead to the formation of outflows that are 
able to cover the inner parts of the visible disc. 
In \citet{superpaper}, we consider the spatial properties of quasar accretion
discs affected by a spherically symmetric 
scattering envelope formed by the wind produced by super-Eddington accretion. 
However, in such a model, the mass accretion rates required seem to be
unexpectedly large, $\mdot> 10^3$, where \mdot\ is mass accretion rate in
dimensionless units,
\begin{equation}\label{E:mdot}
\mdot = \dfrac{\dot{M}c^2}{L_{\rm Edd}} = \dfrac{\dot{M} \varkappa c}{4\pi GM},
\end{equation}
where $L_{\rm Edd} = \dfrac{4\pi GMc}{\varkappa}$ is the Eddington luminosity
and $\varkappa$ is the (presumably, Thomson) opacity of the medium.
The black hole masses inferred are inconsistent with
the virial mass estimates \citep{supererratum}. There are no indications for
mass accretion rates this high even in the brightest AGN while some estimates
around $\mdot \sim 100$ exist (see for example \citealt{collin}). 
For near-Eddington mass accretion rates, the super-Eddington region is
small and the outflows are too fast ($v\sim \mdot^{-1/2}$) and, hence, the
optical depths are too small. 

This actually means that if one wants to reproduce the observed quasar sizes
with the outflow-based model, the outflows either should be bound (and thus
be disqualified as outflows) or possess nearly zero total mechanical energy
(parabolic motion). The latter seems unrealistic because some process is
required to ensure the outflow has the kinetic energy that is exactly equal to its
potential energy. This problem is eliminated if one assumes that an {\it
  inflow} rather than an outflow is responsible for the formation of the
scattering envelope. In this case, nearly parabolic motion is a natural outcome
of energy conservation if the energy is lost at dynamical or
  longer timescales. The infalling matter may scatter or reprocess the
radiation of the disc, thus making the effective size of the disc larger. 


There is another line of evidence for the existence of scattering material around
supermassive black hole accretion discs, which comes from the broad
emission-line regions (BLRs). 
Broad components of emission lines in quasars and other active galactic nuclei
(AGNs) are currently believed to be formed in a clumpy partially ionized gas
in a geometrically thick disc-like structure \citep{BLRD,BLRBon} surrounding
the accretion disc itself. The matter in
BLRs moves at near-virial velocities possessing at the same time large net
orbital momentum that is comparable to Keplerian orbital momentum 
Until recently, the existence and direction (inflows or outflows) of net radial motions in BLR were debated.
However, recent velocity-resolved reverberation mapping results support
the existence of an inflow in BLRs of many AGNs
\citep{doro12,grier13}. 
Besides, the long-lasting problem of blueshifted high-ionization broad
emission lines \citep{gaskell82} is elegantly explained by scattering inside a
converging translucent flow \citep{gaskell09, GG13} existing in the inner
parts of BLRs. 

\begin{figure}
 \centering
\includegraphics[width=\columnwidth]{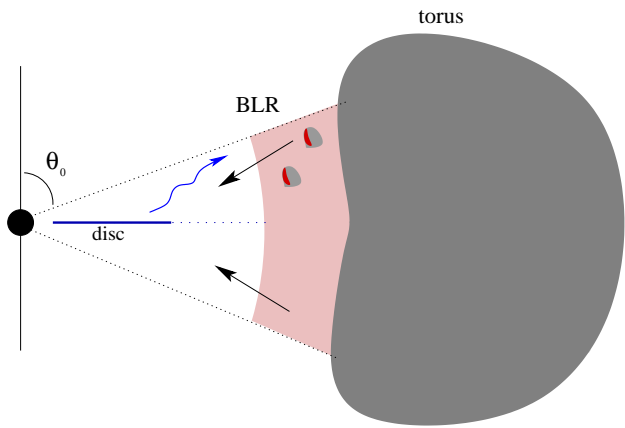}
\caption{ 
Sketch illustrating the unified scheme of AGN accretion combining the dusty
torus, BLR and accretion disc. Radiation from the accretion disc ionizes and
destroys the dusty clouds of the torus and makes them radiate in broad
emission lines. The material from the torus and BLR finally precipitates into
the disc. A couple of BLR clouds are shown schematically with one side
highlighted to indicate the photoionization of the illuminated side of the cloud.
}
\label{fig:dscheme}
\end{figure}

This new observational evidence { allows us to unify } the dusty torus surrounding
most of the well-studied AGN \citep{tristram12}, BLR, and the accretion
disc into a single accretion flow, as shown in Fig.~\ref{fig:dscheme}. 
Transition from the torus towards the
partially ionized gas of the BLR is thought to be connected to dust
sublimation at the temperature of about $T_{\rm dust} \simeq (1\div 2)\times
10^3\rm K  $
, which is weakly dependent on density \citep{dust,nenkova08}. The source
of heating is the radiation of the accretion disc. The broad
lines themselves are emitted in a certain range of radii, generally in between
the size of the dusty torus and the accretion disc. Together with the
evidence for inflow this means that inside certain radius the material emitting in the BLR either
enters the accretion disc or at least becomes too hot to manifest itself via
ultraviolet lines. Both processes probably work and BLR clouds are
gradually heated and destroyed and then precipitate into a thin disc because
of energy loss mechanisms (collisions and subsequent cooling). At the same
time, the geometry of the BLR inherits the oblate geometry of the torus and thus
forms a structure sometimes referred to as a ``bird's nest'' \citep{nests}.

It follows that there should be ionized gas scattering not only the
photons of the broad emission lines but also the radiation of the disc. 
The distance range for this material is dictated,
first, by heating by the disc radiation, and, second, by the cooling
processes that lead to formation of the thin accretion disc. 

In this paper we show that a geometrically thick
scattering inflow with sub-virial radial velocity, which gradually deposits
its material into the accretion disc, in fact can explain the basic features of
the quasar spatial structure resolved by microlensing, such as larger sizes and
weaker dependence on wavelength than predicted by the standard disc
model. 

In Sect.~\ref{sec:str}, 
we construct a model of the ionized inflow. Intensity
distributions for different parameter sets and half-light radii are calculated
in Sect.~\ref{sec:intties}. { In Sect.~\ref{sec:num}, our results are compared
to observational data. } We discuss the results in Sect.~\ref{sec:trep}. 

\section{Accretion flow structure}\label{sec:str}

We assume that the gas accretes at the total rate of $\dot{M}$ from the
gaseous torus in the form of individual optically thick clouds that are
subsequently destroyed by disc radiation and mutual interactions. 
We assume that the radial
velocity (directed inward, taken by absolute value) scales with the local
virial velocity
\begin{equation}
v_r = \beta \sqrt{\dfrac{GM}{R}},
\end{equation}
where $\beta<1$ is a dimensionless coefficient.
The existing net radial velocity
estimates for BLR are sub-virial \citep{GG13} by a factor of several, close to
but smaller than the chaotic velocities observed in BLRs and 
responsible for the virial factor of $f_{\rm v}\sim 5.5$. Virial factor is
  defined as the ratio of the virial velocity {squared
  to the measured velocity dispersion}, $f_{\rm v}=\dfrac{GM}{R \langle v^2\rangle}$ (see
  for example \citealt{park12} and references therein). { 
Random motion velocities of BLR clouds are sub-virial by a factor of
$1/\sqrt{f_{\rm v}} \sim 0.4$ and the radial inflow velocities measured for
Seyfert galaxies \citep{doro12} are a factor of several smaller, which implies
$\beta \sim 0.1$. As we see
    below in Sect.~\ref{sec:num}, the likely value of the relative velocity is
  smaller, of the order $0.01$, but in Sect.~\ref{sec:intties} we adopt $\beta=0.1$.}

We express the mass accretion rate in the disc as $f_{\rm d}(r)
\dot{M}$. All the rest $(1-f_{\rm d})\dot{M}$ is assumed to exist in the form of a
geometrically thick highly inhomogeneous flow
subtending some solid angle $\Omega = 4\pi \cos\theta_0$, where $\theta_0$ is
the half-opening angle of the flow (see Fig.~\ref{fig:dscheme}), which we hereafter consider fixed. 
The density of the geometrically thick flow is written as
\begin{equation}
\rho = \dfrac{(1-f_{\rm d})\dot{M}}{\Omega v_r R^2}.
\end{equation}
This estimate does not
include clumping effects, which are of little importance, as long as
scattering is a linear process in density. In
  appendix~\ref{sec:app:clumpy}, it is shown that scattering opacity becomes
  sensitive to clumping only if the optical depths of individual clumps become
considerably large. 

Normally, BLR clouds are assumed to have hydrogen column
densities of $10^{22}\div 10^{24}\cmsq$,
which makes them optically thin to Thomson scattering. These values are
relevant for ionized gas; the amount of neutral gas is more difficult to
constrain. 
Total line luminosities together with some limit on the total covering
  factor allow us to estimate the thickness of the cloud to be close to
  $10^{23}\cmsq$ (see for instance \citealt{peterson97,peterson06}). 
  However, these estimates are biased by the presence of neutral
  gas and the real column densities may be larger, although there no
  indications that these densities should be larger.
  There are indications \citep{maiolino} for
 the  existence of a population of cometary-shaped clumps that are visible through X-ray
  absorption and probably identical to the population of BLR clouds or
  at least a large part of it. The column densities of these objects are
  estimated as $10^{23}\div 10^{24}\cmsq$. 

In case of optically thick clouds, the effective
optical depth is reduced by about the characteristic optical depth of a
cloud. Here, clumping was taken into account as a correction multiplier for
optical depths $\xi = (1-e^{-\tau_1})/\tau_1 $, where $\tau_1$ is the optical
depth of an individual cloud. Dependence of the observed
intensity distributions upon clumping is also addressed in
Sect.~\ref{sec:radius}. Below, the limit of $\tau_1\to 0$ is used everywhere by
default. 

It is instructive to introduce dimensionless quantities
\begin{equation}\label{E:r}
r = \dfrac{R c^2}{GM},
\end{equation}
and $\mdot$ according to ~(\ref{E:mdot}), and the opacity is set to 
$\varkappa\simeq 0.34\,\rm cm^2 g^{-1}$, which is the Thomson scattering opacity
(for solar metallicity). Then the density becomes
\begin{equation}\label{E:rho}
\rho = \dfrac{4\pi }{\Omega} \dfrac{ (1-f_{\rm d}) \mdot}{\beta}
\dfrac{c^2}{GM\varkappa} r^{-3/2}. 
\end{equation}
The optical depth to scattering in vertical direction is
\begin{equation}\label{E:tauv}
\tau_{\rm v}(r) \simeq \int_0^{R\cos\theta_0} \varkappa \rho dz =
\dfrac{1-f_{\rm d}}{\beta} \dfrac{\mdot}{\sqrt{r}}.
\end{equation}
There is one unknown quantity left in our problem, $f_{\rm d}=f_{\rm
  d}(r)$. We consider that radial and vertical motions both scale with the
virial velocity and the scattering clouds are destroyed while passing through
the equatorial plane. As we show later in Sect.~\ref{sec:press}, 
the column density of the disc is many orders of magnitude
    larger than that of a single cloud, 
therefore BLR clouds should be strongly affected by passage
  through  the accretion disc if the latter has already formed at the radius
  under  consideration. 
The geometrically thick flow is then depleted as
\begin{equation}
\dfrac{d}{dR}\left((1-f_{\rm d}) \dot{M}\right) = 4\pi R \rho v_\theta,
\end{equation}
where the vertical velocity is also considered virial 
$v_\theta = \beta_\theta \sqrt{GM/R}$ with $\beta_\theta = \beta$ (for simplicity).
The partial mass accretion rate itself, according to ~(\ref{E:rho}), equals
$(1-f_{\rm d}) \dot{M} = \Omega R^2 \rho v_r $, hence,
\begin{equation}
\displaystyle \frac{d(1-f_{\rm d})}{dR} =  \dfrac{4\pi}{\Omega} \frac{\beta_\theta}{\beta}
\frac{1-f_{\rm d}}{R} . 
\end{equation}
This implies a power-law dependence on the radius
\begin{equation}
f_{\rm d}(r)=1-\left(\dfrac{R}{R_{\rm out}}\right)^a,
\end{equation}
where $a=\dfrac{\beta_\theta}{\beta}\dfrac{1}{\cos \theta_0} = \dfrac{1}{\cos
  \theta_0}$, $R_{\rm out}$ is the outermost radius where accretion disc
exists{ or, alternatively, the outermost radius where the geometrically
  thick inflow starts to deplete}. 
Outside $R_{\rm out}$, $f_{\rm d}=0$. For simplicity, we assume
here the outer radius of the disc equals the outer radius of the ionized
flow as well. 
The radial optical depth of the scattering flow is then
\begin{equation}\label{E:rtau}
\tau_r = \int_{r_{\rm in}}^r \varkappa \rho(r) dR \simeq
\dfrac{4\pi}{(a-1/2)\Omega} \dfrac{\mdot}{\beta} \dfrac{r^{a-1/2}}{r_{\rm
    out}^a} .
\end{equation}
Here, $r_{\rm in}$ is the inner radius of the disc. 
In our assumptions, $a\geq 1$ and thus the lower integration limit is
  relatively unimportant, especially if $r_{\rm out}\gg r_{\rm in}$. 
The outer radius $R_{\rm out}$ should be a physically
motivated quantity that is connected to the processes that lead to destruction of the
torus clouds and are responsible for radial drift. The actual value of
$R_{\rm out}$ is determined by the radiation fields that destroy dust and ionize
the gas composing the clouds. Ionization is the key process because the
most relevant process is scattering by
free electrons.\footnote{Rayleigh scattering is also expected to be important
  in BLR clouds; see \citet{GG13}. Scattering by dust may be of importance at
  larger distances because dust albedo is still high, about 0.5, in the far
  ultraviolet according to observational data and most popular interstellar
  dust models (see \citet{dust} and references therein).}
Large density contrasts and complex geometry
make it impossible to estimate the outer radius as the size of the
Str\"{o}mgren region. 
The only clear and simple criterion that can be used in this case is the
radiation density temperature of the radiation of the disc.
We assume that the outer radius of the ionized inflow is
determined by some radiation energy density temperature $T_{\rm out}$,
\begin{equation}
\displaystyle \frac{4\sigma_{\rm SB}}{c} T^4_{\rm out} = \frac{L}{4\pi R^2},
\end{equation}
\begin{equation}\label{E:rout}
r_{\rm out} = \displaystyle \sqrt{\frac{c^5 \eta \mdot}{GM \varkappa \sigma_{\rm SB}
    T^4_{\rm out}}} \simeq 10^3 \left(\dfrac{10^4\rm K}{T_{\rm out}}\right)^2
\sqrt{\eta \mdot \frac{10^9\Msun}{M}}.
\end{equation}
Here, $\eta$ is radiative efficiency of the accretion disc,  which we
  later assume is equal to 0.057 in all numerical estimates, as for a
  Schwarzschild black hole. 
This condition is similar to the dust evaporation condition defined by energy density
temperature $T_{\rm ED} =T_{\rm dust} \simeq (1\div 2)\times 10^3\rm K$
\citep{dust}. The proposed limit
$T_{\rm out}$ should be not lower than $T_{\rm dust}$ because dust efficiently
absorbs EUV radiation
and shields the gas from the source. 

Total radial optical depth is calculated using expression~(\ref{E:rtau})
evaluated at $r=r_{\rm out}$,
\begin{equation}\label{E:taur}
\begin{array}{l}
\displaystyle
\tau_r(r_{\rm out}) = \dfrac{4\pi}{(a-1/2) \Omega} \frac{\mdot}{\beta}
\frac{1}{\sqrt{r_{\rm out}}}  \\
\qquad{}\displaystyle \simeq \dfrac{0.03}{(a-1/2) \cos \theta_0}
\frac{\mdot^{3/4}}{\beta \eta^{1/4}} \left(\frac{M}{10^9\Msun}\right)^{1/4}
\frac{T_{\rm out}}{10^4\rm K} .\\
\end{array}
\end{equation}
A fraction of about $\tau_r \Omega/4\pi$ of the disc luminosity is
captured by the inflow and may contribute to the observed radiation
of the disc. This quantity may be considerably high even for
$\mdot\lesssim 1$ depending on $\beta$ and the radial extent of the scattering
region. 

\section{Observed intensity distribution}\label{sec:intties}

We consider the observed intensity distribution composed of two parts: 
directly visible accretion disc radiation and
accretion disc radiation scattered by the inflow. Accretion disc
radiation conforms more or less to the standard disc model \citep{SS73}, but
the mass accretion rate becomes smaller at large radii, $\dot{M}_{\rm d} = f_{\rm d}
\dot{M}$. Besides, at large radii the radiation of the disc is  seen
through the scattering material. The disc is assumed to be observed at the inclination
of $i$, while the effects of inclination are
neglected for the scattered radiation. Below we assume $\cos i =1$ (disc seen face on) if not stated
otherwise. Propagation of the photons scattered by the
outer scattering flow is difficult to treat exactly, hence, we consider that some
fraction of the disc monochromatic luminosity (primarily the radiation
initially going at large inclinations, which have a large probability to be scattered in the inflow) is scrambled and redistributed over
the radial coordinates. Finally, monochromatic intensity may be expressed in
the form,
\begin{equation}\label{E:idist}
I_\nu = I_\nu^{\rm disc} e^{-\tau_{\rm v}\xi}\cos i + (1-e^{-\tau_{\rm v}\xi})\dfrac{f L_\nu}{4\pi^2 R^2},
\end{equation}
where $\tau_{\rm v}$ is the vertical optical depth calculated according
to~(\ref{E:tauv}), $\xi = (1-e^{-\tau_1})/\tau_1$ is clumping correction
factor ($\tau_1$ is the optical depth of a single cloud; see
Appendix~\ref{sec:app:clumpy}),
$f\simeq \cos\theta_0$ is the luminosity fraction scattered
inside the inflow ($f=\cos\theta_0$ for an isotropic source  and
$f=\cos^2\theta_0$ for a flat disc without limb darkening; relativistic
effects also distort the angular distribution of the radiation of the inner
parts of the disc),
$L_\nu$ is the cumulative luminosity of the disc released inside current
radius
\begin{equation}
L_\nu = 4\pi^2 \int_{0}^{R} I_\nu^{\rm disc} R' dR',
\end{equation}
an additional $\pi$ multiplier originating from the transition between
flux and intensity, and the local disc intensity is defined by the standard disc theory as the local
Planck radiation with the temperature determined by the energy release in the
disc, given here according to \citet{SS73} as follows:
\begin{equation}
I_\nu^{\rm disc} = \dfrac{2h\nu^3}{c^2} \dfrac{1}{\exp(h\nu/kT(R))-1},
\end{equation}
\begin{equation}
T(R) \simeq \left(\dfrac{3}{2} \dfrac{c^5}{\sigma_{\rm SB}\varkappa GM}
\mdot \right)^{1/4} \left(1-\sqrt{\dfrac{r_{\rm in}}{r}} \right)^{1/4} \times
r^{-3/4} f_{\rm d}^{1/4} .
\end{equation}
This expression is for a standard, non-relativistic accretion disc with a mass
accretion rate modified by the factor $f_{\rm d}$. If the dimensionless mass
accretion rate becomes more than about several tens, the accretion disc
may become super-Eddington in its inner parts \citep{poutanen}. In this case, the innermost
parts of the disc are covered by another scattering envelope that we
found,  in \citet{supererratum}, to be too small, for reasonably low
accretion rates, to explain the observational data. 

The photon flux from
  the disc as seen by the scattering medium should also be attenuated by a
  factor of $e^{-\tau_r\xi }$. However, the scattered radiation has a large (of
  the order of $f$) probability to be scattered once more. 
 Multiple scatterings
  effectively make the radiation isotropic, eventually reproducing the
  inverse-square scaling with distance, hence we neglect this
  multiplier and assume that the emissivity of the scattered radiation
  decreases simply $\propto R^{-2}$. The real intensity distribution
    should be of course more complicated.

\begin{figure}
 \centering
\includegraphics[width=\columnwidth]{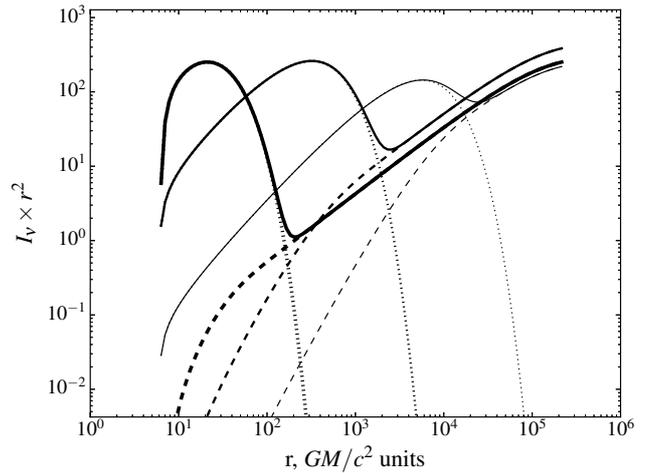}
\caption{ 
Intensity (multiplied by $r^2$, in relative units) as a function of radius
for $\theta_0 = 45^\circ$, $\mdot=100$,
$M_{\rm BH}=10^8\Msun$, $T_{\rm out}=10^3{\rm K}$, and $\beta=0.1$, $i=0$
(disc is seen face-on). Three different
energies were considered, 0.1, 0.3, and 1\Ry\ (from the thinnest to the thickest curves). 
Solid lines show observed (face-on) intensity distributions, dotted
lines correspond to standard disc intensity distribution, and dashed lines to the contribution of
the scattered radiation. 
}
\label{fig:intty100}
\end{figure}

\begin{figure}
 \centering
\includegraphics[width=\columnwidth]{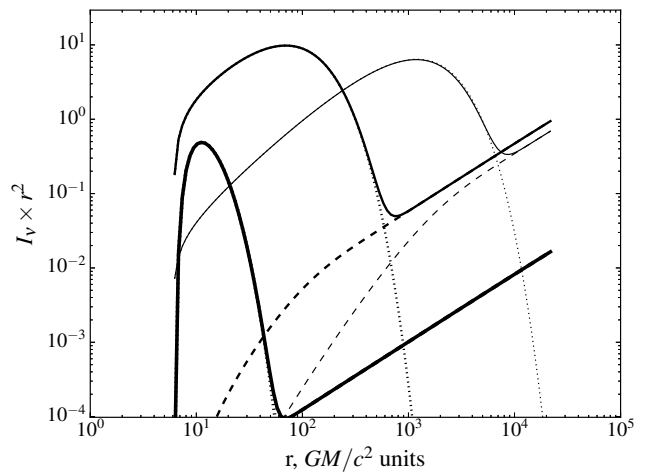}
\caption{ 
Same as previous figure, but for $\mdot=1$. }
\label{fig:intty1}
\end{figure}

Sample intensities are shown in Figs.~\ref{fig:intty100} and
\ref{fig:intty1} for two different mass accretion rates ($\mdot = 1$ and
$\mdot=100$) and three different photon frequencies. 
Depending on the model parameters, spatial properties may be
different, but the two-component structure holds for a very broad range of
mass accretion rates and outer temperatures. 
After introducing the scattering component, radius dependence on wavelength
always becomes shallower. At larger
BH masses, the inner and outer edges of the disc set a limit for disc size
variations that is independent of the scattering component. 
At higher dimensionless mass accretion rates, the optical depth of
the outer flow becomes large and the contribution of the scattered radiation
becomes important. In this latter case, the effective size of the disc increases if the
scattered luminosity becomes comparable to the visible luminosity of the
disc. 

\subsection{Apparent disc radius}\label{sec:radius}

\begin{figure*}
 \centering
\includegraphics[width=1.0\textwidth]{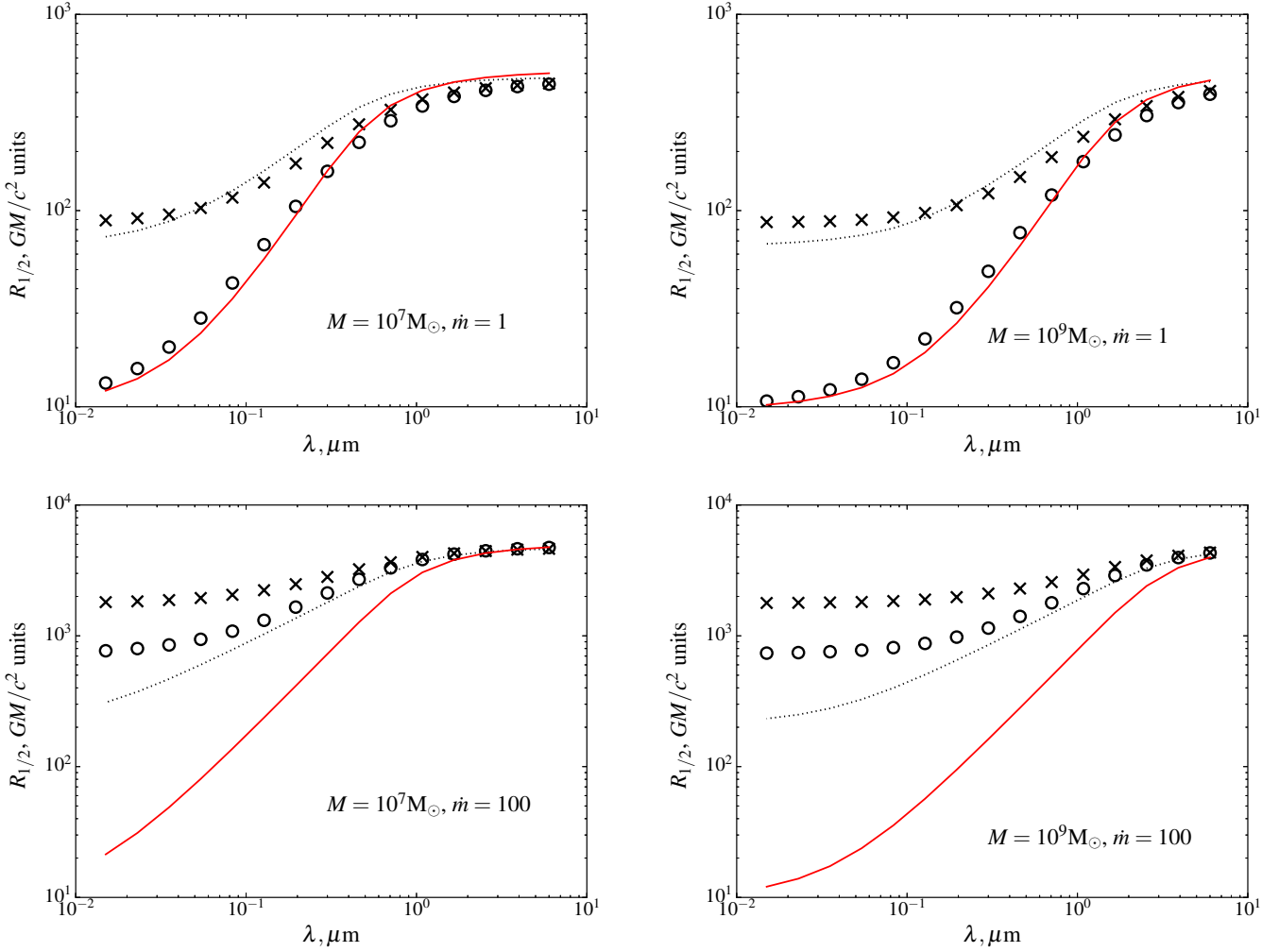}
\caption{ 
Half-light radii (open circles) and intensity-weighted mean radii (crosses) 
for $\theta_0 = 45^\circ$, $T_{\rm out}=5\times 10^3{\rm
  K}$, and $\beta=0.1$. BH masses are $10^7\Msun$ (left panels) and
$10^9\Msun$ (right panels). Dimensionless mass accretion rates are $\mdot =1$
(upper panels) and $100$ (lower panels). Standard accretion disc
predictions are plotted with {red } solid lines, dotted lines show approximation
$R_{1/2}=\sqrt{R_{\rm d} R_{\rm out}}$, where $R_{\rm d}$ is the
  half-light radius of a standard accretion disc. 
}
\label{fig:rmean}
\end{figure*}

\begin{figure*}
 \centering
\includegraphics[width=1.0\textwidth]{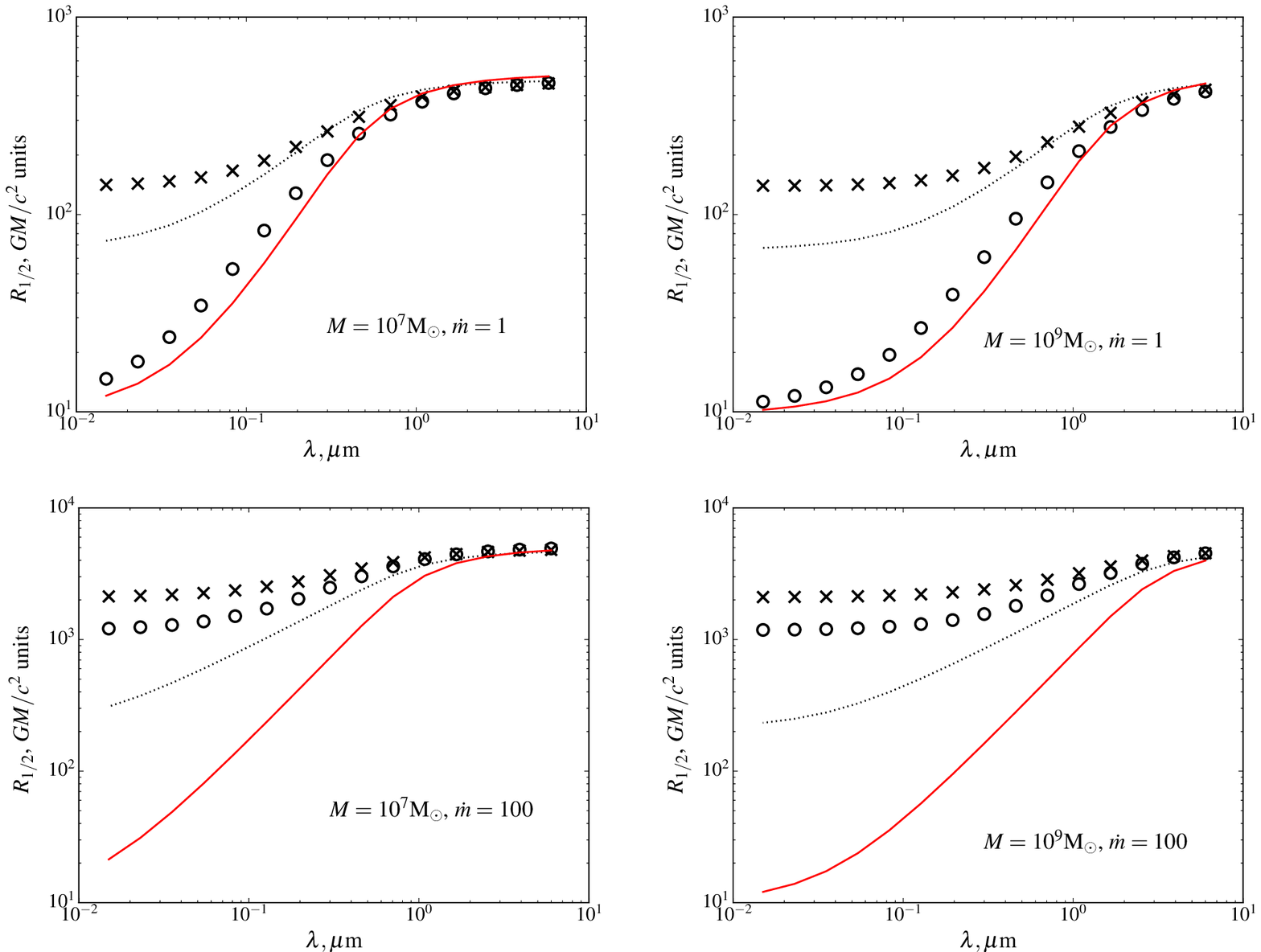}
\caption{ 
Same as previous figure, but for an accretion disc inclined by
$i=60^\circ$. 
}
\label{fig:rcos}
\end{figure*}

The quantity primarily
responsible for microlensing properties of a source is its half-light radius $R_{1/2}$
inside which half of the observed flux is emitted \citep{mortonson},
\begin{equation}\label{E:rhalf}
\displaystyle \frac{1}{2} = \dfrac{\int_0^{R_{1/2}} I R dR}{ \int_0^{+\infty} I R dR}.
\end{equation}
This parameter may be contrasted to another quantity that is more convenient for
analytical estimates and much more sensitive to scattered light contribution,
namely to the intensity-weighted mean radius,
\begin{equation}
\langle R \rangle = \dfrac{\int_0^{+\infty} I R^2 dR}{ \int_0^{+\infty} I R dR}.
\end{equation}
Two quantities may differ severely if the intensity
distribution has extended wings. Half-light radius is actually a
non-linear median estimate, which practically ignores wings in intensity
distribution if their contribution to the flux is small. At the same time, the mean
radius may be affected even by a very faint extended halo. This
is well illustrated by Fig.~\ref{fig:rmean} where the mean radius is
strongly increased by the existence of faint scattering wings while the half-light
radius remains practically unaffected if the mass accretion rate is small
($\mdot \lesssim 1$). In Fig.~\ref{fig:rcos}, identical quantities are
plotted for inclination 
$i=60^\circ$. In this case, accretion disc contribution is two
times smaller and the effective radii are larger for large mass accretion
rates. { We do not consider here the contraction of the
  apparent disc size due to projection effects. }

For the intensity distribution introduced above (Eq. \ref{E:idist}),
one can estimate the effective mean radius in assumption of $\tau_{\rm v}\gg 1$
  and $L_\nu$ independent of distance,
\begin{equation}
\langle R \rangle = \dfrac{\langle R \rangle_0 L_\nu+ f
  \left(R_{\rm out}-R_{\rm in}\right) \times L_\nu}{\left(1+f\right) L_\nu} \simeq
\langle R \rangle_0 + f R_{\rm out},
\end{equation}
where $\langle R \rangle_0$ is the half-light radius of an accretion disc
without a scattering halo. 

At the same time, it may be shown that whenever the half-light radius considerably exceeds
the half-light radius of the accretion disc $R_{\rm d}$, it should scale
approximately as $\sqrt{R_{\rm out} R_{\rm d}}$. Indeed, approximating the
expression for intensity as $I^{\rm disc}_\nu+fL_\nu/R^2$, one can re-write the
definition of the half-light radius (\ref{E:rhalf}) as
\begin{equation}
\displaystyle \int_0^{R_{1/2}}I^{\rm disc}_\nu RdR + f \frac{L_\nu}{4\pi^2}
\ln\left(\dfrac{R_{1/2}}{R_{\rm d}}\right) =  \dfrac{1}{2} \frac{L_\nu}{4\pi^2} \cdot \left(1 + f
\ln\left(\dfrac{R_{\rm out}}{R_{\rm d}}\right) \right).
\end{equation}
The $R_{\rm d}$ limit here arises from the cumulative nature of the luminosity
$L_\nu$ , which makes the scattering term vanish at the radii smaller than
$R_{\rm d}$. Finally, one can re-write this condition as
\begin{equation}
\displaystyle \int_0^{R_{1/2}}I^{\rm disc}_\nu RdR = \dfrac{1}{2} \frac{L_\nu}{4\pi^2} \cdot \left( 1 + f \ln
\left( \dfrac{R_{\rm out} R_{\rm d}}{R_{1/2}^2} \right) \right).
\end{equation}
Because the expression on the left should be larger than
$\int_0^{R_{\rm d}}I^{\rm disc}_\nu RdR =  \dfrac{L_\nu}{8\pi^2}$, the radius $R_{1/2}$ can not
exceed the geometric mean $\sqrt{R_{\rm d} R_{\rm out}}$. If the contribution of the
halo becomes important and intensity behaves as $I_\nu \propto R^{-2}$, the
half-light radius approaches the geometrical mean of some effective inner
radius and the outer radius $R_{\rm out}$,
\begin{equation}
R_{1/2}\simeq \sqrt{R_{\rm d} R_{\rm out}} .
\end{equation}
There are two important outcomes of this estimate. Firstly, the radius
scales with wavelength as $R_{1/2}\propto \lambda^{2/3}$ in this limit. 
Dependence
on wavelength becomes even shallower as the disc radius becomes comparable to
$R_{\rm out}$. Second, $R_{1/2}$ exceeds $R_{\rm d}$ only if $f$ and $R_{\rm out}$ are large
enough. The scattered flux fraction $\sim \cos \theta_0
/\cos i$ depends on the geometrical
thickness of the flow and on the inclination of the disc. 

\begin{figure*}
 \centering
\includegraphics[width=1.05\textwidth]{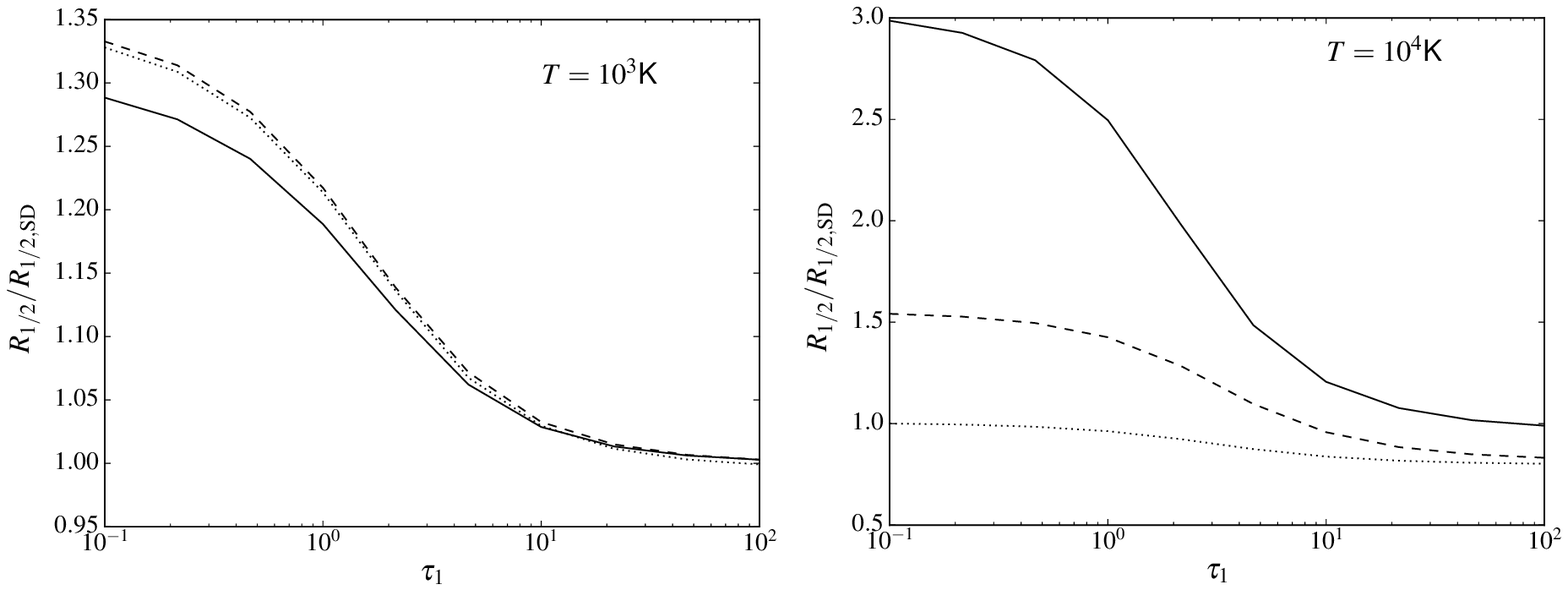}
\caption{ Half-light radius dependence upon the optical depth of a single
  cloud for different wavelengths of 0.1 (solid line), 0.3 (dashed), and 1${\rm
    \mu m}$ (dotted). Radius is expressed in the units of the standard disc
  half-light radius at the same wavelength. Black hole mass
  $M_{\rm BH} = 10^8\Msun$ mass accretion rate $\mdot=10$, temperature limits
  $10^3$ (left), and $10^4$K (right). The disc is seen face-on, $\theta_0=45^\circ$.
}
\label{fig:tvar}
\end{figure*}

Clumping effects make the scattering medium more transparent thus decreasing
the scattered fraction of radiation but increasing the mean radius of the
halo. Sometimes it leads to a non-monotonic dependence of radius correction
factor upon the wavelength. Effects of the optical depth 
of a single cloud $\tau_1$ are shown
in Fig.~\ref{fig:tvar} for two different outer temperatures.  Generally,
increase in $\tau_1$ decreases the contribution of the scattered
radiation, but the effect becomes profound only for $\tau_1 \gg 1$ and scales
inversely proportional to $\tau_1$ for large optical depths. 

\section{Numerical estimates and comparison to observations}\label{sec:num}

We use observational accretion disc sizes and structure parameters to
check the validity of the model and constrain its parameters. 

\subsection{Dependence on wavelength}\label{sec:num:zeta}

\begin{figure*}
 \centering
\includegraphics[width=1.05\textwidth]{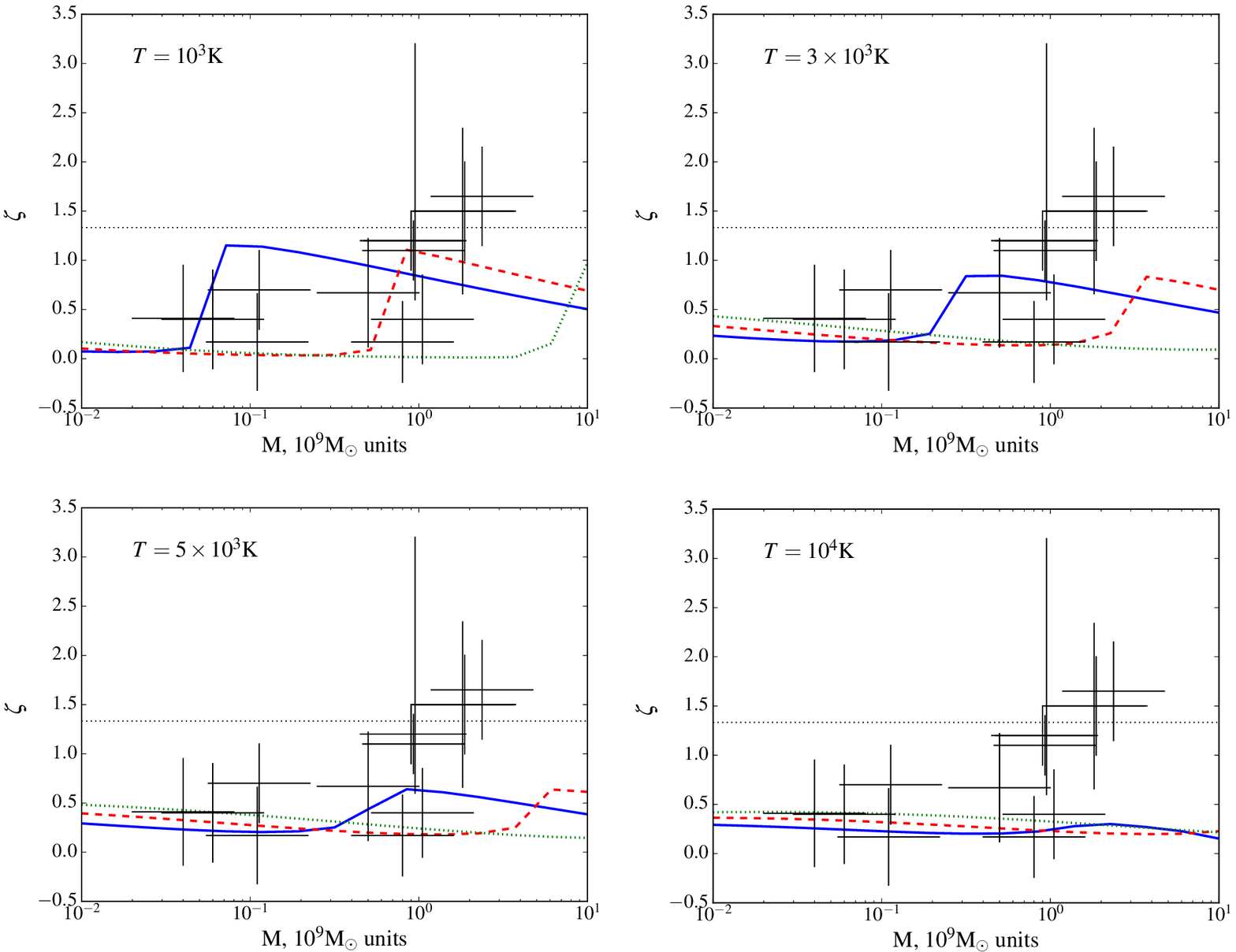}
\caption{ Comparison of structure parameter $\zeta$ estimated from
  observational data (black crosses) and calculated numerically. { Virial
    mass is shown along the x-axes.} From left to
  right, the outer temperature values are $10^3$, $3\times 10^3$ (upper
  panels), $5\times 10^3$, and $10^4$K (lower panels). 
  The three curves in every case correspond to different absolute
  mass accretion rates (solid blue, dashed red, and dotted green curves
  correspond to 1, 10, and 100$\Msunyr$, respectively). { The horizontal dotted line
  corresponds to the standard disc limit $\zeta=4/3$.}
Everywhere the inflow
  half-opening angle is $\theta_0=45^\circ$. 
}
\label{fig:zms}
\end{figure*}

We parameterize the dependence of the half-light radius on wavelength as
$R_{1/2}\propto \lambda^\zeta$ and calculate the values of $\zeta$ in the
relevant (``big blue bump'') wavelength range $0.1\div 1\,\rm \mu m$. Outside this
wavelength range, the accretion disc is likely not the primary radiation
source. Shorter wavelengths may be also affected by the relativistic effects
and super-Eddington accretion regime in the inner disc parts. 
For an infinite standard disc (without any
outer or inner limits), $\zeta =4/3$, but several processes, including those
considered in the current work, make this quantity lower. In
Fig.~\ref{fig:zms}, the estimated values of $\zeta$ are compared to the
values obtained by fitting the
observational data of \citet{blackburne12}. 
We set $\beta=0.01$ because this value allows us to explain the apparent
  accretion disc sizes (see Sect.~\ref{sec:morfit}). In each panel,
  the temperature is fixed and each of the curves corresponds to a fixed physical
  mass accretion rate. 

Several observational radius estimates for different wavelengths between 0.1
and 1$\,\rm \mu m$ were
used to make the power-law fits shown by black crosses. 
The power-law slopes best fitted to the
observational data were presented in \citet{superpaper}. The real shapes of the
$R(\lambda)$ dependences are more complex than a true power law, usually
showing flattenings at higher and lower wavelengths similar to those expected
for an accretion disc model with finite outer and inner edges (see
Figs.~6-8 in \citealt{blackburne12}). The average slopes of
these curves are often close to zero but still positive and, in a couple of
cases, they are very close to the slope of a standard disc. 

The mass accretion rate in the outer parts of the disc
is probably determined by external processes and thus does not depend on the mass of
the black hole. 
If the mass accretion rate in physical units $\dot{M}$ does not
depend on the black hole mass $M$, dimensionless mass accretion rate scales
inversely proportional to the mass $\mdot \propto M^{-1}$, and the effects of
scattering and distorted intensity distribution become more important for
smaller black holes (since the inflow optical depth for a smaller black hole
mass is higher; see
equation~(\ref{E:taur})). In Fig.~\ref{fig:zms}, we attempt to reproduce the $M
- \zeta$ plot with a population of black holes accreting at fixed mass
accretion rate (from 1 to 100\,\Msunyr) for different outer temperatures. 
Using an outer temperature value about $(1\div 3)\times 10^3$K and dimensional
mass
accretion rates in the range $3\div 10\,\Msunyr$ allows us to explain the overall
behaviour of the $\zeta-M$ plot. Lower mass supermassive black holes appear in
a moderately super-Eddington regime ($\mdot \gtrsim 100$, at least several
times larger than the critical mass accretion rate), and for these black holes outflows
can also affect the properties of the disc. However, the size of the envelope
formed by the outflow is expected to be smaller than the scattering region
(see Sect.~\ref{sec:inout} for a more detailed discussion). 

The dimensionless mass accretion rate enters all the
optical depths (equations \ref{E:tauv} and \ref{E:taur}) in combination
$\mdot/\beta$. Hence, for instance, apparently large mass accretion rates may
be mimicked by small radial velocities. 

Strong deviations from the standard disc power-law asymptotic occur when the
flux from the halo becomes comparable to the visible flux from the disc. This
requires a significant percentage of the disc luminosity to be scattered
($f\gtrsim 0.5$, $\mdot/\beta \gtrsim 10$). For the heaviest objects of the sample (see
Fig.~\ref{fig:zms}), observational $\zeta$ seems to marginally exceed the standard disc
value. There are several effects that can reproduce large structural
parameters $\zeta\simeq 2$, one of which is irradiation. If the radiation of
the disc is thermalized instead of scattered, the temperature of
thermalized radiation scales with radius as $T\propto r^{-1/2}$ , which implies
$r\propto \lambda^{2}$. 

 { 
\subsection{Apparent disc sizes for objects with known fluxes}\label{sec:morfit}

Additional constraints may be made by considering the objects where mass
accretion rates may be estimated independently. 
We applied our model to single spectral band accretion disc size
measurements made by \citet{morgan10}. The advantage of these data set is the
existence of de-lensed fluxes that allow us to make independent mass accretion
rate estimates.
We consider the sample from
\citet{morgan10} with the data for one object (\qj) replaced by newer data from
\citet{morgan12}, following our work in \citet{superpaper}. Mass accretion rate was
estimated using monochromatic de-lensed fluxes from \citet{morgan10}. Standard
multi-colour accretion disc approximation allows us to link monochromatic flux
(at observer frame $\lambda_{\rm obs}=0.79{\rm \upmu m}$, which is close to
the observable I-band flux)
with the mass accretion rate assuming the black hole mass is known. Using
expression (9) from \citet{superpaper}, we can estimate the mass accretion
rate as a function of the observed magnitude of the disc as
\begin{equation}\label{E:mdotest}
\mdot \simeq 7.4\times 10^{-4} \left( \frac{10^9\Msun}{M}\right)^2 \left(
\frac{D_{\rm A}}{\Gpc}\right)^3 10^{-0.4(I-19)}\left(1+z\right)^4\cos^{-3/2}i,
\end{equation}
where I is the observed I-band magnitude, $z$ is redshift, and $D_{\rm A}$ is
the angular-size distance (about $1.5\div 1.7$\Gpc\ for all the objects). 

Objects and their basic properties are listed
in Table~\ref{tab:morgan}. Half-light radii calculated using our model were
compared to the observational results using a crude $\chi^2$ criterion,
\begin{equation}
 \displaystyle \chi^2 = \sum \left(\frac{ \lg{R_{1/2}}-\lg{R_{1/2\mbox{,
      model}}}}{\Delta\lg{R_{1/2}}}\right)^2,
\end{equation}
where the uncertainty $\Delta\lg{R_{1/2}}$ is half-width of 70\% interval in
$\lg{R_{1/2}}$ measured by
microlensing effects. Summation was performed for the whole set of 11 objects.
We assumed $i=60^\circ$ and $\theta_0 =
60^\circ$. In Fig.~\ref{fig:tbfit}, contours of constant $\chi^2$ are
overplotted with the shades representing the predicted values of $\zeta$ in
the spectral range $0.1\div 1{\rm \upmu m}$.

\begin{table*}\centering
\caption{ Object set used for fitting in Sect.~\ref{sec:morfit}.  
 Redshifts $z$, virial masses $M_{\rm vir}$, half-light radii $R_{1/2}$,
de-lensed I-band magnitudes $I_{\rm corr}$, dimensionless mass accretion rates,
and observed structure parameters $\zeta_{\rm obs}$ are given. 
} 
\label{tab:morgan}
\bigskip
\begin{tabular}{c c c c c c c c c}
\hline
\hline
Object ID& $z$  & $M_{\rm vir}$, $10^9\Msun$ & $R_{1/2}$,
  $10^{15}$cm & $I_{\rm corr}$ & $\mdot$ & $\zeta_{\rm obs}$\\
\hline
QJ 01584325 & 1.29 & 0.16 & 4.9$\div$19.4 & 19.09$\pm$0.12 & 2.3$\div$50.5 &
-- \\ 
HE 04351223 & 1.689 & 0.5 & 2.4$\div$38.7 & 20.76$\pm$0.25 & 0.0$\div$1.2 &
0.1$\div$1.1 \\ 
SDSS 0924+0219 & 1.52297 & 0.11 & 1.0$\div$4.9 & 21.24$\pm$0.25 &
0.3$\div$10.1 &  0.15$\div$0.8 \\ 
FBQ 0951+2635 & 1.24603 & 0.89 & 12.2$\div$77.2 & 17.16$\pm$0.11 &
1.0$\div$21.1 & -- \\ 
SDSS 1004+4112 & 1.73995 & 0.39 & 1.0$\div$3.9 & 20.97$\pm$0.22 &
0.1$\div$1.6&-- \\ 
HE 11041805 & 2.3192 & 2.37 & 9.7$\div$30.7 & 18.17$\pm$0.31 & 0.1$\div$4.5 &
1.65$\pm$0.5 \\ 
PG 1115+080 & 1.73547 & 1.23 & 38.7$\div$193.8 & 19.52$\pm$0.27 & 0.0$\div$1.2 &
0.4$\pm$0.5 \\ 
RXJ 11311231 & 0.654 & 0.06 & 3.1$\div$7.7 & 20.73$\pm$0.4 & 0.2$\div$8.3 &
0.4$\pm$0.5 \\ 
SDSS 1138+0314 & 2.44375 & 0.04 & 0.5$\div$7.7 & 21.97$\pm$0.19 & 2.9$\div$79.0 &
0.4$\pm$0.5  \\ 
SBS 1520+530 & 1.855 & 0.88 & 7.7$\div$19.4 & 18.92$\pm$0.13 & 0.2$\div$5.3 &
--\\ 
Q 2237+030 & 1.695& 0.9 & 4.9$\div$19.4 & 17.9$\pm$0.44 & 0.5$\div$25.7 &
1.15$\pm$0.2 \\ 
\hline
\end{tabular}
\tablefoot{ Except for Q J01584325, all the
  radii were estimated in \citet{morgan10}, data for Q J01584325 were taken
  from \citet{morgan12}. Mass accretion rates were estimated using the
  de-lensed flux values of \citet{morgan10} in assumption of isotropic
  emission using formula~\ref{E:mdotest}. Structure parameter estimates $\zeta_{\rm obs}$ were
  made by us \citep{superpaper} using the data by \citet{blackburne12}.}
\end{table*}

\begin{figure}
 \centering
\includegraphics[width=1.0\columnwidth]{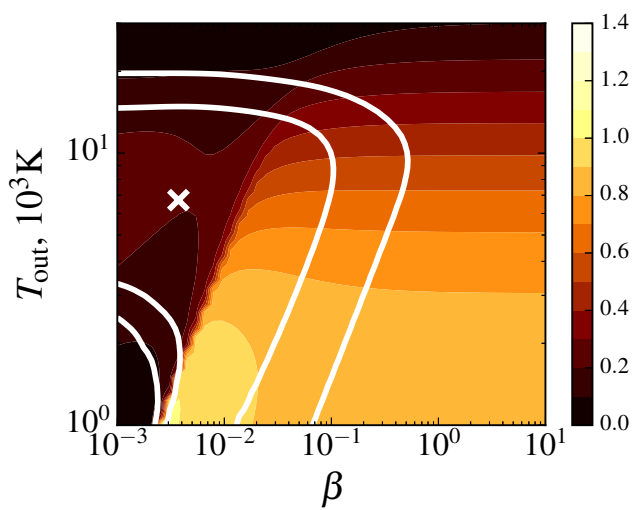}
\caption{ Structure parameter $\zeta$ for a $M=10^9\Msun$, $\mdot=1$ disc as
  a function of $\beta$ and outer temperature. Overplotted are lines of constant
  $\chi^2$ (minimal value $+2.3$ and $+4.6$, corresponding to 70\% and 90\%
  probability if two parameters are varied) for the
  fitting of disc sizes of the \citet{morgan10} sample. 
}
\label{fig:tbfit}
\end{figure}

The fluxes allow us to estimate the mass accretion rates of $\mdot \sim 0.1\div
10$ for the sample of objects, which makes an upper limit for $\beta$.
As it is shown in Fig.~\ref{fig:tbfit}, observational data favour $\beta
\lesssim 0.1$ and set an upper limit for $T\lesssim 2\times 10^4$K. 
The minimal value of $\chi^2$ is about 1.5 for 9 degrees of freedom that means
that the model is over-defined for the existing data, and more comprehensive
statistical analysis is required in future. However, it also means that
observed accretion disc sizes are easily reproduced by a model like ours. 

The values of $\beta$ favoured by our fitting are much smaller than the
  characteristic velocity dispersions in virial units ($\sim 1/\sqrt{f_{\rm
      v}}\simeq 0.4$) known from broad line profiles. 
Small radial
velocity may result from the inefficient angular momentum transport in the
inflowing matter. To move inward, the scattering clouds should have some means
of angular momentum loss or redistribution. { For the case of accretion discs,
this is provided by turbulent or magnetic viscosity.}
We consider several processes
possibly responsible for angular momentum transfer in the outer, geometrically
thick, part of the flow in Sect.~\ref{sec:trep}.

Lower temperatures
$T_{\rm out} \lesssim 3\times 10^{3}{\rm K}$ make it possible to reproduce
different $\zeta$ for different objects. As can be seen in
Fig.~\ref{fig:zms}, high temperatures restrict the range of possible
structure parameters, hence the existence of nearly standard discs in some quasars
is naturally reproduced only for $\beta \sim 0.01$, $T_{\rm out}\sim (1\div
3)\times 10^{3}{\rm K}$. The existence of two populations of quasars, with small
and large (about unity) $\zeta$ , is then easily explained by roughly universal
physical mass accretion rates about $\dot{M}\sim 3\div 30\Msunyr$ (see
Fig.~\ref{fig:zms}). 

All these results were obtained for $\theta_0 =60^\circ$. 
Smaller solid angles require smaller $\beta$, approximately as $\beta \propto
\Omega$. It is difficult to put any constraints with the data on hand,
but $\Omega\ll 1$ would make it impossible for the scattered component to
provide an important contribution. Probability of secondary scatterings is likely
to go to zero in this case hence expression~(\ref{E:idist}) is no longer valid,
and the secondary (scattered) intensity is limited by the value $\sim
\dfrac{\Omega}{4\pi} I^{\rm disc}_\nu$. Therefore, we argue that the solid
angle should be large, probably $\dfrac{\Omega}{4\pi} \gtrsim \dfrac{1}{2}$,
for our model to work. 

}

\section{Discussion}\label{sec:trep}

{ 
\subsection{Opening angle and covering factor}\label{sec:CF}

For our model, it is important that the scattered flux is comparable or
exceeds the primary flux from the accretion disc. This requires a large
covering factor of the scatterer, $\dfrac{\Omega}{4\pi} \gtrsim
\dfrac{1}{2}$. 

The geometry of dusty tori was studied for different kinds of AGN, mainly obscured
Seyfert galaxies, in several papers \citep{ibarlira07,ichikawa15}. The estimated
half-opening angles are about 60$^\circ$. On the other hand,  about 50$\div$60\%  of   AGN are obscured \citep{davies15}, which again results in
$\theta_0 \sim 50\div 60^\circ$. 

The vertical structure of a BLR, however, does not show any single
characteristic opening angle. For instance, \citet{BLRvert} study the
geometry of the region in different lines for several nearby well-studied
Seyfert galaxies. The relative thickness of the region $H/R$ varies from $\sim
0.1\div 0.3$ for Balmer lines to $\sim 1$ for hotter lines such as
\ion{C}{iv}$\lambda$1549. Although the uncertainties are still very high and
the effect is less studied at higher luminosities, it suggests the solid angle
of the region is large, possibly larger than the solid angle of the dusty
torus. The possible inward increase of the solid angle of the BLR does
  not necessarily contradict the proposed settling of the material into the
  disc. If the formation of the disc is driven by cloud collisions, as  proposed in Sect.~\ref{sec:press}, some of the matter may be easily brought
  to orbits with larger inclinations. Besides, radiation and wind from the
  accretion disc can affect the motion of the clouds through drag and outward
  pressure. Details of the physics and the
  interpretation of the observational
  results are uncertain.

If the radiation of the disc is scattered efficiently, it is geometrically
channelled in a solid angle $4\pi-\Omega$, which means that the observed
scattered flux is enhanced by the factor of
$\sim \left(\dfrac{4\pi}{\Omega}-1\right)^{-1}$. This becomes important for
$\theta \lesssim \pi/3$, when the amplification factor is about two. We
did not consider this effect as it relies on the unknown beaming pattern of
the radiation of the disc and exact geometry of the scattering region. 

\subsection{Mass accretion rates}\label{sec:mdot}

As we have already shown, the half-light radius starts deviating from that of
the disc if the scattered flux becomes comparable to the flux from the
disc. The scattered fraction is roughly $\dfrac{\Omega}{4\pi} \tau_r$, where
$\tau_r$ is given by (\ref{E:taur}), and becomes around unity when
dimensionless mass accretion rate reaches the critical value
\begin{equation}
\displaystyle \mdot_{\rm cr} \simeq 100
\left(a-\frac{1}{2}\right)^{4/3}\left(\beta\frac{10^4{\rm K}}{T_{\rm out}}\right)^{4/3}
\eta^{1/3} \left(\frac{M}{10^9\Msun}\right)^{-1/3}.
\end{equation}
For realistic values of $\beta \sim 0.01$, $T\simeq 10^3{\rm K}$,
$a-\dfrac{1}{2}\simeq 1$, $\eta \simeq 0.06$, and $M\sim 10^9\Msun$ the critical mass accretion
rate $ \mdot_{\rm cr} \simeq 2$. 
More detailed calculations presented in Sect.~\ref{sec:num} support this
estimate. Adjusting the free parameters $\beta$, $T_{\rm out}$, and $a$ (or the
solid angle of the flow, as $a\simeq \cos^{-1}\theta_0$) allows us to shift the
critical value in broad limits, where the most probable value is around
unity. However, large $\mdot$ cannot be excluded either. 

Evidently, $\mdot \gtrsim 100$ produces a luminosity of $\gtrsim 100\eta \sim 10$ in
Eddington units, which makes the accretion disc supercritical in its inner
parts. The region of the disc where the local Eddington limit is violated is
restricted by the spherization radius $r_{\rm sph}\simeq
\dfrac{3}{2}\mdot$ \citep{SS73, poutanen}. On the other hand, the monochromatic half-light radius
of the standard accretion disc is, according to
\citet{morgan10}, about $2.44$ radial scale lengths, or
\begin{equation}
\displaystyle r_{1/2} \simeq 170 \left(\frac{\lambda}{1\mu}\right)^{4/3} \mdot^{1/3} \left(\frac{M}{10^9\Msun}\right)^{-1/3},
\end{equation}
which is either close to or larger than the spherization radius for the
parameters we are interested in. Therefore, the effects of disc thickness,
advection, and outflows are important mainly for the EUV part of the spectrum
$\lambda \lesssim 0.1{\rm \upmu m}$ that is still poorly covered by microlensing
studies. However, if the supercritical part of the disc forms an outflow, the
pseudo-photosphere of the outflow may become comparable in size { to} the accretion
disc. { We considered this pseudo-photosphere in
\citet{superpaper, supererratum} and conclude that its contribution can affect
} the apparent disc size at
shorter wavelengths $\lambda \sim 0.1{\rm \upmu m}$. However, as we show below in Sect.~
\ref{sec:inout}, the effect of an inflow is generally stronger than that of a
supercritical wind. 

}

\subsection{Accretion disc formation and interaction with the BLR
  clouds}\label{sec:press}

In our model, the material is supposed to enter the accretion disc through
destruction of BLR clouds. This statement needs some justification. For
instance, if the surface mass density of a cloud is much larger than that of
the accretion disc, it easily passes through the disc losing only a small
portion of its vertical momentum. On the other hand, collision with a high
surface density disc leads to significant loss in vertical momentum. Besides, such a collision
is highly supersonic and probably leads to destruction of the cloud. 

Column number densities of BLR clouds are usually estimated as $N_{\rm BLR} \sim
10^{23}\div 10^{24}\cmsq$ although most of the estimates only constrain the
amount of ionized gas and are thus only lower limits
\citep{KG00,KG15}. The surface column density of the
accretion disc, on the other hand, may be estimated using the equations of the
standard accretion disc theory \citep{SS73} as
\begin{equation}\label{E:discNH}
\small
\displaystyle N_{\rm disc} \simeq 
\footnotesize
\left\{ 
\begin{array}{lc}
\displaystyle 7.2\times 10^{30} \alpha^{-4/5} \mdot^{3/5}
\left(\frac{M}{10^9\Msun}\right)^{1/5} r^{-3/5}\cmsq, &
r\lesssim 1200\mdot^{2/3},\\
\displaystyle 2.6\times 10^{31} \alpha^{-4/5} \mdot^{7/10}
\left(\frac{M}{10^9\Msun}\right)^{1/5} r^{-3/4}\cmsq, &
r\gtrsim 1200\mdot^{2/3} .\\
\end{array}
\right.
\end{equation}
The two cases correspond to the so-called zones b and c of the standard disc,
differing in dominating opacity sources. The inner {radiation-pressure
dominated} zone a is too compact compared to the size of the BLR. 
The dimensionless radius $r$ here should be close to the outer radius that we
consider in our model or to the empirical size of the BLR that is known to
scale approximately as $R_{\rm BLR} \propto \sqrt{L}$ \citep{kaspi07}. The
relation given by \citet{kaspi07} may be written as 
\begin{equation}\label{E:BLRrad}
\displaystyle R_{\rm BLR} \simeq 0.2 \sqrt{\eta \mdot \frac{M}{10^9\Msun}} \pc,
\end{equation}
if one identifies the UV luminosity with the
bolometric luminosity $\lambda L_\lambda(1350\AAA) \simeq L = \eta \mdot
L_{\rm Edd}$. Here, \mdot\ is the dimensionless mass accretion rate
(introduced above, see Eq.~(\ref{E:mdot})) and $\eta$ is the overall
  accretion efficiency $\eta = \dfrac{L}{\dot{M} c^2}$. The product $\eta
  \mdot = L/L_{\rm Edd}$ is also known as the Eddington factor. 

In terms of dimensionless radius, 
\begin{equation}\label{E:BLRrelrad}
\displaystyle r_{\rm BLR} \simeq 4000 \sqrt{\eta \mdot\frac{10^9\Msun}{M} },
\end{equation}
close to the outer radius given by expression~(\ref{E:rout}) for $T_{\rm
  out}\simeq 2000{\rm K}$. 

At this radius, the 
hydrogen column density in the disc $N_{\rm disc} \sim 10^{28}\div
10^{30}\cmsq$, greatly exceeding the column densities estimated for BLR
clouds. Unless all the column densities in BLR are vastly underestimated, the
accretion disc easily becomes an impassible barrier for BLR clouds and thus
efficiently absorbs their material. 

Formation of the seed accretion disc that
subsequently drains mass from the BLR may be connected to cloud-cloud
interaction. 
Indeed, for a given spherical cloud of density $n$ and column density $N$,
the number of collisions per second is about
\begin{equation}\label{E:nucc}
\nu_{\rm cc}\simeq  \frac{\pi}{4} \left(\dfrac{N}{n}\right)^2 v n_{\rm c},
\end{equation}
where $n_{\rm c}$ is the volume density of the clouds, and $\displaystyle
v\sim \sqrt{\frac{2GM}{f_{\rm v} R}}$ is the mean relative collision
velocity. The number density of the clouds may be expressed through mass accretion
rate, radial velocity, and cloud mass as
\begin{equation}\label{E:nclouds}
n_{\rm c} \simeq \dfrac{\dot{M}}{4\pi R^2 v_r M_{\rm c}}, 
\end{equation}
where $M_{\rm c} \simeq N^3 n^{-2} m_{\rm p} $ is the mass of a cloud.
Neglecting multipliers of the order unity, 
\begin{equation}\label{E:nucc1}
\nu_{\rm cc} \simeq \dfrac{GM\mdot }{\beta \sigma_T c R^2 N }.
\end{equation}
At the empirical BLR radius given by the estimate~(\ref{E:BLRrad}), this
frequency becomes
\begin{equation}\label{E:nucc:est}
\nu_{\rm cc} \sim 10^{-9} \frac{1}{\beta} \frac{M}{10^9\Msun} \dfrac{10^{23}\cmsq}{N} {\rm s}^{-1},
\end{equation}
that is close to or even higher than the local Keplerian frequency
$\Omega_{\rm K} = \sqrt{GM/R^3}\sim 10^{-9} (M/10^9\Msun)^{-1/4} (\eta \mdot)^{-3/4}
{\rm s}^{-1}$  meaning that cloud collision is an important process that
  may be driving the radial flow inside the BLR and formation of the accretion
  disc. 

\subsection{Optical depths of inflows and outflows}\label{sec:inout}

Whenever a geometrically thick inflow exists, its optical depth should be
higher than that of the corresponding outflow for two reasons: (i) the
mass ejection rate cannot exceed the mass inflow rate and (ii) the velocity of
the outflow should be larger. The second condition requires quantitative
estimates. If one considers wind accelerated in a gravitational field of a
point mass $M$ at some distance $R$, most of its velocity is gained near the
sonic surface (or some sonic surface analogue if non-thermal processes are
responsible for acceleration; see \citealt{winds}), where the velocity of the
wind should exceed the local parabolic velocity. Otherwise, the flow remains
bound and falls back into the accretion disc; this possibility is also
interesting as a possible explanation for the spatial properties of quasar
discs. Non-zero terminal velocity of the wind is the result of kinetic energy
exceeding the potential energy at the sonic radius. At larger distances, it is
practically constant and
\begin{equation}
v_{\rm w} = \beta_{\rm w} \sqrt{\dfrac{GM}{R_{\rm s}}},
\end{equation}
where $\beta_{\rm w}$ is some dimensionless constant of order unity and {$R_{\rm s}$ is
the sonic radius. }
 We note that $\beta_{\rm w} \ll 1$ requires fine tuning between the kinetic and
potential energy in the wind acceleration region.  
At the same time, depending on its angular momentum, the radial
velocity of an inflow may be close to or much smaller than the virial
\begin{equation}
v_{\rm in} = \beta \sqrt{\dfrac{GM}{R}},
\end{equation}
where $\beta$ can be significantly smaller than unity.
Difference in velocities implies difference in optical depths by
a factor of $\dfrac{\tau_{\rm in}}{\tau_{\rm w}} \sim \dfrac{\beta_{\rm w}}{\beta}
\dfrac{\dot{M}}{\dot{M}_{\rm w}} > 1$. While the optical depths may be
comparable, the
spatial distributions of the scattering matter are different. If one considers
a spherically symmetric density distribution, its optical depth from some
radius $R$ to infinity is
\begin{equation}
\tau(R) = \varkappa \int_r^{+\infty} \rho dR.
\end{equation}
Pseudo-photosphere radius $R_1$ is defined by condition $\tau(R) =1$. If the
velocity is parabolic everywhere, 
\begin{equation}
\displaystyle \rho(R) = \frac{\dot{M}}{4\pi R^2 v(R)} = \dfrac{c^2}{GM\varkappa} \frac{\mdot}{\beta} r^{-3/2},
\end{equation}
which implies $\tau = 2\mdot /\beta \sqrt{r}$ and dimensionless radius of the
pseudo-photosphere $\displaystyle r_1 = 4\left(\frac{\mdot}{\beta}\right)^2$.
This may be contrasted to the super-virial outflow case that we consider in~\citet{superpaper}.
Optical depth reaches unity for $r_1\sim \mdot^{3/2}$ for a hyperbolic outflow and
for $r_1\sim \mdot^2$ for a parabolic inflow. This allows scattering inflows
to produce more extended halos and pseudo-photospheres for identical mass
accretion rates as long as they stay sub-virial. 

\subsection{Dynamics of the inflow}\label{sec:trep:dyn}

If a significant portion of the radiation of the disc is scattered by a
geometrically thick inflow, one can also expect this radiation to have some
dynamical effect upon the scattering medium. However, independent of the mass
accretion rate, this effect is insufficient to power the angular momentum loss
inside the scattering region. Tangential momentum flux carried out by
scattered radiation may be calculated as the momentum
flux density of the radiation $\dfrac{L}{4\pi c R^2}$ multiplied by the local
tangential velocity in light velocity units $v_\varphi/c$. Hence the radiation
acts upon the scattering medium with the torque
\begin{equation}
K \simeq \dfrac{\Omega}{4\pi} L_{\rm disc} \cdot v_{\varphi} R/c^2 \sim
\dfrac{\Omega}{4\pi} \eta \dot{M} \sqrt{GMR}.
\end{equation}
Here, $\eta$ is the radiative efficiency of the disc and  $\Omega$ is the
  solid angle subtended by the inflow. 
On the other hand, the angular momentum flux carried by the accreting matter 
through the same surface of
radius $R$ is $\dot{M} \sqrt{GMR} > K$ for any combination of parameters since
$\eta < 1$, $\Omega < 4\pi$. Radiation drag is at least several
times short in powering the angular momentum transfer in the inflow. 

\subsection{Self-consistent mass accretion rate}

While radiation of the disc is unable to power angular momentum transfer, 
this radiation can induce angular momentum transfer by magnetic fields by keeping the
matter ionized. If the number of
hydrogen-ionizing quanta generated by the disc is $Q_{\rm H}$, the ionized mass is
$M_{\rm g} \simeq m Q_{\rm H} / \alpha_{\rm r} n_{\rm e}$. For $Q_{\rm H} \sim 10^{56} \rm s^{-1}$,
hydrogen recombination coefficient $\alpha_{\rm r}\simeq 2\times 10^{13}
\rm cm^3
s^{-1}$, mean particle mass $0.6$ of a proton mass ($m\simeq 0.6m_{\rm p}$),
and $n_{\rm e} \sim 10^{11}\cmc$ (about or somewhat lower than the densities typically
estimated for BLRs; see for example~\citet{negrete12}),
the ionized mass is about several solar masses. The estimated BLR masses are
much higher owing to incomplete ionization \citep{BLRmass}. If all the ionized
material moves inward with sub-virial velocities as was assumed here, its
mass infall rate is
\begin{equation}
\dot{M}_{\rm in} \simeq \beta M_{\rm g} \Omega_{\rm K}(R_{\rm out}),
\end{equation}
where $\Omega_{\rm K}(R) = \sqrt{GM/R^3}$ is the local Keplerian frequency.
Substituting here the above expression for the mass of the ionized gas and
expression~(\ref{E:rout}) for the outer radius, one obtains
\begin{equation}
\dot{M}_{\rm in} \simeq \dfrac{\beta m Q_{\rm H}}{\alpha_{\rm r} n_{\rm e}}
\sqrt{GM}\left(\dfrac{\varkappa \sigma_{\rm SB} T_{\rm out}^4}{\eta \mdot GMc}\right)^{3/4}.
\end{equation}
If the inner disc temperature is hotter than one Rydberg (Ry), the ionizing
quanta
production rate $Q_{\rm H}$ grows with the mass accretion rate
approximately as $Q_{\rm H}\sim L / {\rm Ry} = \eta \dot{M} c^2/{\rm Ry}$. Some part of this value,
namely $\Omega/4\pi$, is intercepted by the inflow. Assuming
$\dot{M}_{\rm in}=\dot{M}$, one can obtain an estimate for
self-consistent mass accretion rate as follows:
\begin{equation}\label{E:mrate}
\begin{array}{l}
\displaystyle \mdot \simeq \dfrac{\eta^{1/3}\varkappa \sigma_{\rm SB}
  T_{\rm out}^4}{c\left(GM\right)^{1/3}} \left(\dfrac{\Omega\beta}{4\pi}
\dfrac{m c^2}{ \Ry \alpha_{\rm r} n_{\rm e}}\right)^{4/3}
 \simeq \\
\displaystyle \simeq 7\times 10^{-4} \eta^{1/3}
\left(\frac{\Omega}{4\pi} \frac{\beta}{0.01} \frac{10^{11}\cmc}{ n_{\rm e}}\right)^{4/3}
\left(\frac{M}{10^9\Msun} \right)^{-1/3} \left(\dfrac{T_{\rm out}}{10^3{\rm K}}\right)^4.\\
\end{array}
\end{equation}
{ This estimate means that, for a broad range of parameters allowed by the
  fitting and estimates made in Sect.~\ref{sec:num}, the radiation is
  insufficient to ionized significant part of the flow. }
An alternative way to start angular momentum transfer at the regions as far as
$R_{\rm out}$ is through cloud collisions that may happen considerably often,
as shown in Sect.~\ref{sec:press}. 

Partial ionization of the accretion flow also means that pure scattering may
not be as important as reprocession of radiation, as already mentioned
in Sect.~\ref{sec:num:zeta}. The existence of a preferred radius at which the
radiation is reprocessed (close to $R_{\rm out}$ or the radius of the BLR)
suppresses the proposed increase in $\zeta$. The measured ``disc'' radius
should instead approach the value of $R_{\rm out}$ at the wavelengths where
reprocessed emission dominates. The contribution of reprocessed emission was
already proposed to be important for explaining the shape of the quasar ``big
blue bump'' in the UV-to-optical spectral range and may be
equally important for understanding its spatial properties \citep{lawrence12}. 

\section{Conclusions}

Different sites may be responsible for scattering the radiation of the discs
around supermassive black holes. At super-Eddington mass accretion rates, the
inner parts of the disc are expected to be covered by the pseudo-photosphere
of the disc wind. At the same time, near the outer limits of the disc,
partially ionized material associated with BLR or some parts of it may
produce a scattering halo. It seems that this effect is even more important
for the measured half-light radii. Such a model can explain the observed
shallow dependences of the quasar half-light radii on wavelength and the
generally larger observed disc sizes. The required mass accretion rates may
be close to Eddington, but may be much lower if the inflow velocity is
less than or about several per cent of virial. The size of the scattering region should
be close to the distance where energy density temperature reaches $T_{\rm out
}\sim (1\div 3)\times 10^3$K, that is close to dust evaporation limit. 
The covering factor of the inflow should be
comparable to unity ($\cos\theta_0 \gtrsim 0.5$) to make the contribution of
the scattered emission measurable. 

\begin{acknowledgements}

Research was supported by Academy of Finland grant 268740 and by 
Russian Scientific Foundation (RSF) grant 14-12-00146 (clumping corrections
to opacity).
The author would like to thank Juri Poutanen and Anna Chashkina for
discussions. 

\end{acknowledgements}

\bibliographystyle{aa}
\bibliography{mybib}

\appendix

\section{Radiation extinction by clumpy medium}\label{sec:app:clumpy}

The outcome of small-scale, but strong inhomogeneities may be illustrated by
the following toy model. We consider scattering medium consisting
of small identical clouds, each having optical depth $\tau_1$. We also
assume the clouds distributed
randomly in space. The number of clouds $x$ encountered by a given light beam on some
distance $l$ should have Poissonian distribution
\begin{equation}\label{E:app:Poisson}
P(x) = \mu^x e^{-\mu}/ x!,
\end{equation}
where $\mu =  n_{\rm c} A l$ is the expectation for $x$,  $n_{\rm c}$ is
the number density of the
clouds, and $A$ is the cross-section of a cloud. If one considers the clouds
to be spheres of a single radius $R_{\rm c}$, $\mu = n_{\rm c} \pi R_{\rm
  c}^2 l =
\dfrac{3}{4} \dfrac{F}{R_{\rm c}} l$, where $F$ is the volume fraction (filling
factor) of the clouds. Because $\tau_1 \simeq \dfrac{1}{F}\sigma n R_{\rm c}$,
{ where $\sigma$ is the extinction cross-section, and $n$ is the mean density
  (hence the density inside a cloud is $n/F$), }
the mean optical depth is independent of filling factor,
\begin{equation}\label{E:app:meantau}
\langle \tau \rangle = \tau_1 \mu = \dfrac{3}{4} \dfrac{F l}{ R_{\rm
    c}} \tau_1 \simeq \sigma n l.
\end{equation}
Up to the factor of order unity, this quantity equals the optical depth of an
unclumped medium with identical mean density. 

At the same time, the optical depth becomes a stochastic variable because of
the randomly distributed number of clouds encountered. Dispersion of the
number of clouds equals $\mu$ due to the properties of Poissonian
distribution, hence for $\mu\lesssim 1$ variations in optical depth are
important. Because we are primarily interested in attenuated intensity, it is
instructive to estimate the mean value of intensity attenuation factor,
\begin{equation}\label{E:app:dint}
\displaystyle \langle e^{-\tau}\rangle = \sum_{x=0}^{+\infty} \frac{e^{-\tau_1
x}e^{-\mu} \mu^x}{x!} = e^{-\mu}   \sum_{x=0}^{+\infty}
\frac{\left(e^{-\tau_1} \mu\right)^x}{x!} = e^{-\left(1-e^{-\tau_1}\right)\mu}.
\end{equation}
 Thus, extinction effectively becomes lower 
with respect to the case
of uniform medium. One can express the above
result in terms of effective optical depth
\begin{equation}\label{E:app:teff}
\tau_{\rm eff} = \dfrac{1-e^{-\tau_1}}{\tau_1} \langle \tau\rangle.
\end{equation}
Effective optical depth depends only upon the optical depth of a single
cloud and evolves from $\langle\tau\rangle$ at small $\tau_1$ to $\sim
\langle\tau\rangle/\tau_1$ for high
$\tau_1$; there is no intrinsic dependence upon the filling factor or cloud size { Smaller effective optical depth is a consequence of
the increasingly large probability that a particular line of sight will not hit any
cloud. 

\label{lastpage}

\end{document}